\documentclass[conference]{IEEEtran}
\usepackage[marginal,hang]{footmisc}

\usepackage{multirow}
\usepackage{subfigure}
\usepackage{cite}
\usepackage{colortbl}
\usepackage{hhline}
\usepackage{flushend}
\usepackage{color}
\usepackage[bf,small]{caption}
\usepackage{indentfirst}
\usepackage{url}
\usepackage{subfigure}
\usepackage{tikz}
\usepackage{bbding}
\usepackage[all]{xy}
\usepackage{array}
\usepackage{multirow}
\usepackage{multicol}
\usepackage{balance}
\usepackage{etex}
\usepackage{grffile}
\usepackage{paralist}
\usepackage[ruled,noend]{algorithm2e}
\usepackage{graphicx}
\usepackage{booktabs}
\usepackage{color}
\definecolor{darkblue}{RGB}{47,99,99}

\newcommand{\descr}[1]{\vspace{0.2cm} \noindent \textbf{#1}}

\begin{document}
\pagenumbering{arabic}
\thispagestyle{plain}
\title{{\em ``They brought in the horrible key ring thing!''} Analysing the Usability of Two-Factor Authentication in UK Online Banking}
\color{black}

\author{\IEEEauthorblockN{Kat Krol,
Eleni Philippou,
Emiliano De Cristofaro,
M. Angela Sasse}
\IEEEauthorblockA{University College London\\
E-mail: \{kat.krol.10,eleni.philippou.13,e.decristofaro,a.sasse\}@ucl.ac.uk}
}


\maketitle
\begin{abstract}
To prevent password breaches and guessing attacks, banks increasingly turn to two-factor authentication (2FA), requiring users to present at least one more factor, such as a one-time password generated by a hardware token or received via SMS, besides a password. We can expect some solutions -- especially those adding a token -- to create extra work for users, but little research has investigated usability, user acceptance, and perceived security of deployed 2FA.
This paper presents an in-depth study of 2FA usability with 21 UK online banking customers, 16 of whom had accounts with more than one bank. We collected a rich set of qualitative and quantitative data through two rounds of semi-structured interviews, and an authentication diary over an average of 11 days. Our participants reported a wide range of usability issues, especially with the use of hardware tokens, showing that the mental and physical workload involved shapes how they use online banking. Key targets for improvements are (i) the reduction in the number of authentication steps, and (ii) removing features that do not add any security but negatively affect the user experience. 
\end{abstract}

\section{Introduction}\label{sec:introduction}
Online banking provides users with anywhere-anytime access to financial services, enabling customers to check their balance and their recent transactions, pay bills, or issue bank transfers, without having to visit a bank branch.
Over the past decade, an increasing number of banks worldwide began to support online banking, 
with consequent fast-growing adoption rates. In 2013, it was estimated that 42\% of EU and 51\% of US customers use online banking -- a remarkable increase from five years earlier~\cite{eurostat,pew}.

Naturally, remote access to financial services prompts the need for strong security guarantees, including strong authentication. To toughen the security of password-based authentication,
online banking operators increasingly turn to two-factor authentication (2FA), requiring users to provide one or more additional authentication factors -- such as a one-time passcode generated by a hardware token or received via SMS -- besides their password.
In recent years, regulation of several countries has started to mandate its use in the financial sector, though in different forms~\cite{burr_electronic_2013,council2005}.

The need for an extra authentication factor  often imposes additional burden onto the users, as 2FA might require them to carry additional devices with them, increase the time needed to authenticate as well as the number of mistakes and lockouts. As shown in prior work~\cite{adams1999users}, %
since security is not the primary task, users are often frustrated by complex authentication tasks. For instance, enforcing long passwords, frequent changes, and preventing re-use across accounts, often drives them to find the simplest password that barely complies with the requirements~\cite{adams1999users}. Also, imposing lengthy authentication tasks often takes a toll on productivity~\cite{strouble2009productivity} and causes severe disruption to the user~\cite{inglesant2010true}.
Nonetheless, little research has focused on studying (or improving) the usability of 2FA, thus motivating the need to elicit and analyse insight on issues faced by regular users as well as their coping strategies. 

\descr{Overview.} 
In this paper, we present an in-depth study of 2FA usability in the context of UK online banking ($N=21$). We focus on  banking as it represents one of 2FA's most popular contexts of use, and recruit actual customers that have been using 2FA. We conduct two series of face-to-face interviews and ask participants to maintain an authentication diary (over an average of 11 days). We analyse data both qualitatively and quantitatively and uncover a wide range of issues experienced by users -- in particular, with the use of hardware tokens.
We find that participants' satisfaction is negatively correlated with the use of hardware tokens as well as
the need to provide multiple credentials.
The demands placed on users, such as the need to produce and remember a wide range of different credentials (often with confusing nomenclature), carrying and operating extra devices, are quite disruptive to their primary task, and shape how often, from where, and how, customers use online banking. For instance, one participant even switched banks to avoid using a hardware token.

\descr{Contributions.} 
Our study derives important insight on actual authentication interactions, as we interview actual users of 2FA and online banking---unlike prior work, where study participants only performed artificial tasks on mock-up prototypes (cf. Section~\ref{sec:rw}).  %
By analysing the data collected from semi-structured interviews (Section~\ref{sec:results}) and authentication diaries (Section~\ref{sec:diary}), we shed light on struggles and misconceptions experienced by users. 
Our work provides an exploratory analysis of these issues and paves the way for further research, including larger hypothesis-driven studies as well cross-analysis of qualitative and empirical data measuring authentication tasks' completion time, mistakes, lockouts, stress and frustration. As discussed in Section~\ref{sec:conclusion}, we highlight a few actionable items for online banking operators, as we highlight the need to reduce the number of authentication steps, minimise confusion, as well as to remove features that do not add any real security but negatively affect user experience.

\section{Related work}\label{sec:rw}

We now review prior work on 2FA usability and online banking.

\descr{2FA usability.} Braz et al.~\cite{braz2006security} were among the first to investigate the usability of 2FA and suggested that, by increasing redundancy, 2FA strengthens security but reduces usability. However,  they did not conduct any actual user study. Strouble et al.~\cite{strouble2009productivity} analysed the effect on productivity of the Common Access Card (CAC), a smart card and photo ID used by US Department of Defense employees. They found that employees were often locked out as they left the card in the reader and almost stopped answering emails from home, concluding that the DoD lost more than \$10M worth of time. Gunson et al.~\cite{gunson2011user} investigated authentication usability in the context of automated telephone banking, asking 62 participants to rate their experience via a 22-item questionnaire: when second factors of authentication were enforced, users felt more secure than when using only passwords or PINs, but at the expense of usability.

Bonneau et al.~\cite{bonneau2012quest} evaluated several password-replacement schemes, and compared security and usability of passwords, OAuth and OpenID~\cite{recordon2006openid}, hardware and phone-based tokens. They did not conduct a user study, but relied on a set of 25 subjective metrics, eight of which intended to capture usability, concluding that 2FA is more secure but less usable than passwords.
 
\descr{Comparative studies of 2FA.} While studies above compared 2FA with simple password-based authentication, another line of work presented comparative studies within 2FA. Weir et al.~\cite{weir2008user} investigated usability perceptions of three 2FA technologies (a push-button token, a card-activated token, and a PIN-activated token). They asked 50 participants to use each option, identify what they liked/disliked, how they would improve the technologies, and answer a short usability questionnaire. Users were then asked to authenticate again with their preferred technology and fill out the usability questionnaire once more. Authors found a strong correlation between preference and perceived convenience, but only a weak correlation between perceived security and perceived convenience/usability.
In a follow-up study, Weir et al.~\cite{weir2010usable} compared the usability of password authentication and two 2FA technologies, specifically, one-time passwords generated by a token or received via SMS. They performed a lab study involving 141 participants and a 30-question survey, and concluded that familiarity with a technology -- rather than perceived usability -- affected user willingness to use that technology. %
The participants in Weir et al.'s studies interacted with prototypes, performing artificial tasks -- they did not log in to their own accounts and they were not interacting with the authentication interface under real-life constraints (e.g., time pressure). In contrast, we interview actual users of online banking services and collect data on actual interactions, thus deriving important insight on problems faced by regular users as well as on their coping strategies.

De Cristofaro et al.~\cite{de2014comparative} presented an exploratory quantitative analysis comparing the usability of three different 2FA technologies, based on a survey completed by 219 US users recruited on Mechanical Turk. They found that users were relatively pleased with 2FA usability, regardless of the specific technology, and regardless of motivations or context of use. However, the perceived usability was somewhat correlated with demographics---e.g., female and older people felt they expended more cognitive effort to use 2FA. Also, in contrast to previous studies, users' perception of trustworthiness was not negatively correlated with ease of use and required cognitive efforts. 
Compared to~\cite{de2014comparative}, our study does not aim to quantitatively compare across technology and does not rely on surveys, rather, it provides an in-depth analysis focused on struggles and misconceptions experienced by actual 2FA/online banking users.
We do so by means of semi-structured interviews, which provide a flexible way of exploring the user's perspective -- especially in the context of  banking authentication, where it is hard to obtain empirical data from the system owners (such as number of failed logins and password resets) who regard the information as commercially sensitive.
Also note that the online banking 2FA landscape in the US is significantly different from the one in UK, as the majority of US banks (at least at the time research in~\cite{de2014comparative} was conducted) only implemented dual credentials, but not 2FA.

\descr{Online banking.} Just and Aspinall~\cite{just2012security} analysed the use of dual credential authentication for online banking from both security and usability points of view. They considered granularity and time of feedback given to users during the authentication steps as main usability properties, and found that some banks delayed feedback by not providing it at screen change, or provided granular feedback too late in the authentication process. They concluded that these issues are likely to confuse users, but did not conduct an actual user study.
Our work complements~\cite{just2012security} as we conduct an in-depth user study aiming to understand authentication for online banking from the users' point of view. Also, while~\cite{just2012security} looked at dual credentials (e.g., two passwords, two PINs, or two challenge questions), we focus on actual 2FA.

\descr{Authentication diaries.} Other studies have also used authentication diaries to understand authentication habits and usability issues. Inglesant and Sasse~\cite{inglesant2010true} introduced password diaries to capture the details of authentication events happening in the wild and found that frequent password changes are perceived as troublesome, that users do not change passwords unless forced to, and that it is difficult for them to create memorable passwords adhering to the policy. Hayashi and Hong~\cite{hayashi2011diary} also analysed two-week diaries to derive the average number of passwords, frequency of use, recall strategies, etc., across users.
Finally, Steves et al.~\cite{steves2014report} asked 23 employees of a US government organisation to keep a diary for every authentication event in one day, and later interviewed them on their authentication experience: the study highlights users' frustration with authentication processes that disrupt their primary task and hinder their productivity, and uncovers a number of coping strategies aimed to minimise the negative impact of security on employees' work. Authors also found that the requirement to use 2FA in the form of a SecureID token made employees log in remotely less often than they would normally do.

\descr{Summary.} Prior work on 2FA usability presented
expert assessments, survey-based studies, and experiments on prototypes, %
incurring a number of shortcomings. Expert assessments did not involve users, yielding findings that only rely on researchers' judgement and often without the benefit of a structured usability assessment technique, such as GOMS, heuristic evaluation or cognitive walkthrough~\cite{mack1994usability}. Survey-based studies asked participants to make hypothetical choices or report behaviours based on what they can remember. Finally, studies with prototypes were performed in the absence of real-life constraints: without reference to a primary task -- such as paying a bill –- or context of use -– paying a bill from your office during lunch break or in a hotel room while traveling.
This highlights the lack of studies focused on actual users of 2FA and online banking, which is crucial to 
understanding how customers use 2FA different technologies for online banking and how these fit into their every-day activities. 

\section{Background}\label{sec:background}
\begin{table}[t]
\centering
\footnotesize
\setlength{\tabcolsep}{.4em}
\begin{tabular}{|l|l|l|l|}
\hline
  {\bf Bank} & {\bf Identification} &{\bf Pre-assigned?} & {\bf Type} \\ 
 \hline
 Barclays & Surname~+~[Membership, &  Assigned & (Varies) \\
 & Card, or Account No] & & \\ 
 \hline  
  Halifax, & User ID & Assigned but & 9+ letters\\
  Lloyds &  & changeable & and/or digits \hspace{-0.15cm} \\
  \hline
  HSBC, & User ID & Chosen & 5+ letters\\
  First Direct & & &and digits\\
 \hline
 Nationwide & Customer Number & Assigned & 10 digits \\
 \hline
 NatWest, RBS,  & Customer Number (DoB &   Assigned & 10 digits  \\
 Ulster &   ddmmyy + 4 digits) & & \\
 \hline  
 Santander & Personal ID &  Assigned & 5+ digits \\
 \hline
 The Co-op Bank & Account No or & Assigned & (Varies) \\
  & Credit Card No &  & \\
 \hline
\end{tabular} 
\caption{Identification mechanisms for online banking with different UK banks.}\label{tab:identification}
\end{table}

\begin{table}[t]
\centering
\footnotesize
\setlength{\tabcolsep}{.5em}
\begin{tabular}{|l|l|l|}
 \hline
  {\bf Bank} & {\bf 1st Factor} & {\bf 2nd Factor}\\ 
 \hline
 Barclays-1 & 8-digit OTP (PINsentry & -- \\
 & or Mobile PINsentry) &  \\
 \hline
 Barclays-2 & 5-digit passcode & 2 of 8+ characters \\
 & & memorable word\\
 \hline
 Halifax, & 8+ character password & 3 of 6+ characters \\
 Lloyds &  & memorable information\\
 \hline
 HSBC, & memorable answer  & 6-digit OTP (Secure Key    \\
 First Direct &  & or Digital Secure Key) \\
 \hline
 Nationwide-1 & memorable data & 3 of 6 digits passnumber \\
 \hline
 Nationwide-2 & 8-digit OTP (card reader) & -- \\
    \hline
 NatWest, RBS, & 3 of 4-digit PIN & 3 of 6+ characters \\
 Ulster & &  password\\
 \hline
 Santander-1 & 6+ character & 5-digit registration \\
 & password & number\\
 \hline
 Santander-2 & 3 of 6+ character & 3 of 5-digit registration\\
 & password & number\\
 \hline
\end{tabular} 
\caption{Authentication mechanisms for online banking with different UK banks.}\label{tab:auth}
\end{table}

This section summarises the two-factor authentication (2FA) landscape for online and mobile banking in the United Kingdom. All information provided is as of summer 2014. We overview authentication mechanisms for login purposes, setting up new payees, and using mobile banking functionality for 11 popular UK banks: Barclays, Co-operative, First Direct, Halifax, HSBC, Lloyds, Nationwide, NatWest, RBS, Santander, Ulster. We focus on these banks as (i)~all participants in our study have an account with at least one of them, and (ii)~they cover the vast majority of the UK market.

All 11 banks in our sample follow a {\em two-step} authentication process to login on the online banking websites. 
In the first step, users provide an {\em identifier} and, only with Lloyds/Halifax, one authentication factor. Then, if the first step is successful, they are asked to authenticate by providing {\em two authentication factors} (one with Halifax and Lloyds). 
In Table~\ref{tab:identification} and~\ref{tab:auth}, we summarise respectively, the different identification and authentication mechanisms used by different UK banks, using banks' original terminology.

During the identification step, banks usually provide users with a \textit{remember me} option: if selected, users do not have to re-enter their identifier when logging in from the same device and browser, unless they clear the cookies. Also note that, in the second step, some banks display to the user a previously selected picture and/or text, aiming to provide some visual confirmation that the user is not on a phishing site.

\begin{figure}[ttt]
\centering
\subfigure[Secure Key\newline (HSBC)]{
  \includegraphics[width=.25\linewidth]{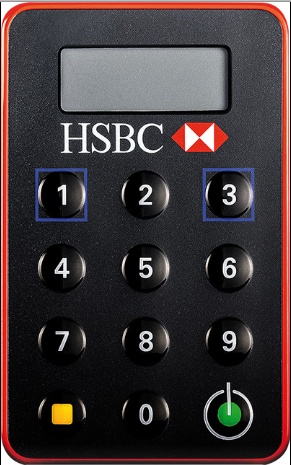}
  \label{fig:hsbc} 
}
\subfigure[PINsentry\newline (Barclays)]{
  \includegraphics[width=.25\linewidth]{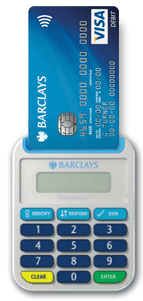}
  \label{fig:pinsentry} 
}
\subfigure[Card-Reader\newline (NatWest)]{
  \includegraphics[width=.25\linewidth]{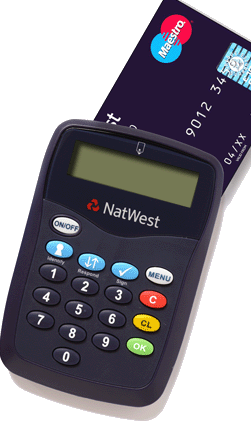}
  \label{fig:natwest} 
}
\caption{Examples of three hardware tokens, i.e., devices generating one-time passwords (OTPs) for authentication purposes.}
\vspace{0.2cm}
\label{fig:all}
\end{figure}

Many of the banks rely on \textit{hardware tokens} to generate the second authentication factor, specifically, one-time passwords (OTPs). Some devices require the user to insert their ``chip-and-PIN'' debit card and enter the card's PIN (e.g., Barclay's PINsentry or NatWest's Card-Reader), while with other devices (HSBC and First Direct's Secure Key) users only need to enter a previously selected PIN. Figure~\ref{fig:all} presents three examples of hardware tokens used by popular UK banks.

When setting up a new payment (e.g., to pay a bill or issue a money transfer to an account that has never paid into), all banks require an additional authentication step (besides those needed for login). Most banks require a one-time password (OTP) authenticating the specific transaction, amount, and/or payee (depending on the bank). 
In Table~\ref{tab:payee}, we report the different technologies used to generate OTPs across different banks.
In the rightmost column, we also introduce some labels to categorise each type of technology. %

All banks in our sample also support mobile banking, via smartphone apps (on iPhone and Android), allowing their customers to check their balance and issue payments. In Table~\ref{tab:mobile}, we report the authentication mechanisms enforced by each bank's app. (We do not report on payment/transaction authentication, but only app login).

\begin{table}[t]
\centering
\footnotesize
\setlength{\tabcolsep}{.6em}
\begin{tabular}{|l|l|c|}
\hline
 {\bf OTP Source} & {\bf Bank} &{\bf Category}  \\ 
\hline
Secure Key & HSBC, First Direct & \textbf{SK}\\
\hline
Card reader & Barclays, Nationwide, NatWest, & \textbf{CR}\\
 &           RBS, Ulster  &\\
\hline
SMS & Santander & \textbf{SMS}\\
\hline
Mobile phone app & Barclays, HSBC, First Direct & \textbf{M}\\
\hline
Phone call & Lloyds, Halifax & \textbf{P}\\
\hline
\end{tabular}

\caption{Technologies used to generate OTPs for different UK banks.
Rightmost column introduces labels, used throughout the paper, to indicate each type of technology.}\label{tab:payee}
\vspace{0.15cm}
\end{table}

\begin{table}[t]
\centering
\footnotesize
\setlength{\tabcolsep}{.6em}
\begin{tabular}{|l|l|l|l|}
\hline
 {\bf Bank} & {\bf ID always} & {\bf Authentication Factor(s)} \\ 
 & {\bf required?} & \\
\hline
Barclays & No & 5-digit PIN\\
\hline
Halifax, & No & 3 characters of\\
Lloyds & & memorable information\\
\hline
HSBC, & Yes & Memorable answer + 3  \\
First Direct & No & characters of a password\\
\hline
Nationwide & No & 3 of 6-digit passnumber \\
\hline
NatWest & No & 5-8 digit PIN\\
\hline
RBS & No & 5-8 digit PIN\\
\hline
Santander & Yes & 3 characters of password,\\
& & 3 digits of security number \\
\hline
Ulster & No & 5-8 digit PIN\\
\hline

\end{tabular} 
\caption{Authentication for mobile apps with different UK banks.}\label{tab:mobile}
\end{table}

\section{Study Design}\label{sec:methodology}
We now present the methodology and the design of our study, consisting of three stages: preliminary open-ended interviews, authentication diaries, and final follow-up interviews. 

\subsection{Methodology}

\descr{Recruitment.} The research was conducted in August 2014 and framed as a user study on online banking. Participants were recruited through social media and from a research participant pool at our university.
Interested users were invited to complete a short pre-screening survey, aiming to assess their eligibility to participate. We collected basic demographic information (age, gender, level of education) and asked potential participants whether or not they had previously used online banking and with which banks. 
Out of 66 people who responded, we selected 21 participants as described below. %

\descr{Sampling and participant demographics.} In order to select a representative sample, we divided potential participants into five groups, one for each category of OTP generator, that is: Secure Key, card reader, one-time password (OTP) delivered via SMS, OTP generated by a smartphone app, and phone-based authentication. All answers given by users were checked for plausibility and cross-referenced later during the interviews. We selected participants to achieve a gender/age balance within each category. 

Our final sample consisted of 11 female and 10 male participants. Mean age was 32.4 (range: 19-69, $SD=10.87$). Out of the 21 participants, 10 of them had a postgraduate degree (a Masters or a PhD), 7 an undergraduate degree, 2 had A-Levels (UK school leaving certification), 1 had some undergraduate education with no completed degree, and 1 participant had completed vocational training. A total of 5 participants had an account with one bank, 14 with two banks, and 2 with three banks. Also, 8 were customers of HSBC, 6  of Barclays, 6 of NatWest, 5 of Santander, 4 of Lloyds, 3 of Halifax, and 2 of RBS. First Direct, Nationwide, the Co-operative Bank, Ulster Bank, and Citigroup all had one customer each.
Finally, 43\% of participants used Secure Key, 76\% card readers, 24\% OTP via SMS, 14\% OTP via smartphone app, and 21\% authentication through a phone call.

\descr{Ethics.} The study design and protocol, along with all materials we used, was reviewed and approved by UCL's Research Ethics Committee (approval number: 6010/001).
Note that we did not collect any personally identifiable information, such as login information or bank details. We temporarily stored participants' demographic and contact information, authentication diaries, and
interview recordings for the duration of the analysis. Both electronic data and diaries were securely disposed of at the end of the study.

\subsection{Study stages}

\descr{Preliminary interviews.} The first stage of the study consisted of semi-structured interviews, conducted in a lab setting with one researcher and one participant at a time. Interviews lasted 30 minutes on average, and involved a total of 21 participants.
The goal of these interviews was twofold: (1)~eliciting a better understanding of banking preferences (e.g.,~``online'' vs. ``offline''), attitudes towards online banking, frequency of use of online and mobile banking services, and (2)~exploring attitudes towards authentication technologies deployed by the banks, aiming to identify sources of common errors, misconceptions, and frustration.
Interviews were structured around a basic set of open-ended questions allowing users to talk about their authentication habits and experience. When needed, the researcher followed up on a topic of interest raised by the participant with further questions (e.g., when one participant identified some part of the authentication process as unnecessarily difficult or unjustifiable).
At the end, participants were introduced to, and briefed on, the next stage -- the diary study and %
received \pounds5 for their participation.

\descr{Authentication diaries.} We asked participants to maintain an banking authentication diary over a 10--12 day period starting immediately after the interviews.
They each received a 20-page booklet (one page for one authentication event). There, participants were asked to write down date/time of the authentication event, name of the bank, the location and the device from where they authenticated, what authentication factor(s) they used, and their reason to authenticate, following some pre-filled fields. Also, they could report on any problems they encountered (and if/how they were solved), they were asked to give a rating of their satisfaction with the authentication process and finally, there was space for any other comments.

\descr{Final interviews.}
The last stage of the study involved 17 participants (80\% of our initial sample), as four participants were unavailable to come for an interview within our set time frame. 
Once again, we conducted semi-structured interviews -- lasting 30 minutes on average -- in a lab setting with one researcher and one participant at a time.
We started with a debrief of the diary entries, asking participants to comment on the experience of keeping an authentication diary, on their authentication routine, and to comment on some of their diary entries. Next, we asked them to talk about other authentication systems they were familiar with, besides the context of online banking, placing more emphasis on systems that they enjoy interacting with, allowing participants to convey what elements they consider vital for a successful authentication process. Finally, we explored participants' perceptions regarding the notion of implicit authentication by explaining the concept and giving examples of how such a system would work.
At the end of the session, participants received \pounds10 for their participation in this final part of the study.

\section{Interview results}\label{sec:results}

In this section, we present the results of a thematic analysis~\cite{braun2006using} 
conducted on both sets of interviews (preliminary and final interviews) by one researcher. The next section will then analyse the authentication diaries.

\subsection{Struggling with hardware tokens} 
One of the most problematic issues highlighted by participants was undoubtedly related to hardware tokens, i.e., devices used to generate one-time passwords (OTPs), such as HSBC's Secure Key, Barclays' PINsentry, NatWest's Card-Reader (see Figure~\ref{fig:all}).

When talking about their experience with hardware tokens, participants discussed both
what they perceived as advantages and disadvantages (e.g., compared to other 2FA forms), with negative feedback being predominant. 
Only four out of 20 participants found that the hardware token was {\em ``easy to use''}, {\em ``portable''}, or {\em ``easy to incorporate into everyday life.''} 
Whereas, seven participants complained that they have to remember to bring it with them, while four found the use of a hardware token to be frustrating. Some found it superfluous (2~mentions), unconvincing (2), not thought-through (2), and five of them questioned the necessity of introducing it. 
We singled out one participant, P19 (SK,CR)\footnote{For each participant we quote, we also indicate, in brackets, the OTP generation technology they use, as per Table~\ref{tab:payee}.}, whose insight somewhat summarises the view of other participants as well:
\begin{quote}
\textit{``I don't think it was really thought-through. I kind of wonder who they consulted in the beginning. [\ldots] It's impractical to be constantly having that on you because if you lost your bag/keys/purse it would be very difficult to then log on or ask for another one. [\ldots] It feels a bit long-winded at times, especially if I am in a rush or I forgot that I had to pay someone. So I find myself in the situation where I often have to order new card readers because I always tend to lose them and then find them. At the moment I am trying to keep one in my bag and one at home. I've probably have about 10 or something because I keep losing them and finding them.''}
\end{quote}

Other participants felt that their account was secure enough before the introduction of the hardware token, as they had never experienced any fraud and/or felt that their password was hard to guess anyway.
P19~(SK,CR) found the online banking authentication procedures to be excessive compared to offline transactions where only a four-digit PIN is needed to authorise a card payment (and, actually, no PIN is required for contactless payments under \pounds20).
Participants often disliked the extra step that the use of the hardware token introduced to their authentication task and found it irritating (3) or inconvenient (5). This seemed to be tied in with the fact that only 1 out of these 8 participants reported using the hardware token to issue bank transfers, which probably contributed to the feeling of superfluousness.

Five participants stressed they thought a particular bank was more secure because it required using hardware token more often (e.g., not only for payment setup but also for login). Four participants (including two from the previous category) said they preferred a bank that used the token less, or did not use it at all. In fact, three participants reported 
changing banks or logging into their account less frequently because of the hardware token. P19~(SK,CR) said:
\begin{quote}
\textit{``Initially, I had an account with HSBC [\ldots] and that was before they brought in the horrible key ring thing that you have to carry with you at all times to log in [\ldots]. With NatWest you have to remember passwords and stuff but you don't have to bring anything physically with you unless you're changing like payments and transfers [\ldots] That's why I'm with NatWest, just because I find their online banking system [\ldots] compared to HSBC, it's the most simple.''} 
\end{quote}

P19~(SK,CR) was also ironic as to why the token was introduced in the first place:
\begin{quote}
\textit{``I've been using Internet 15 years now, and I've used it for a long time. I think people older than me should be using it as well and kind of be more afraid of it\ldots and been told by their banks online banking is more convenient but then found the whole process not really secure. So I think it was about trust issues. It was the idea of having something more tangible similar to the machines in branch or a cash machine.''} 
\end{quote}

P11~(CR,P) reported that they stopped checking their balance and recent transaction while at work, since the login procedure makes it obvious to co-workers they are not working: 
\begin{quote}
\textit{``It's OK when I am at home, but when you are at work and you are pretending you are actually doing work when you are actually checking on your account, then you have to bring out this calculator thing and it's kind of obvious you are not doing work. I'd rather have something where I am just on the screen and it's lot quicker.''}
\end{quote}

\subsection{Struggling with credentials}

\descr{Authentication terminology.} Another important issue relates to confusing terminology used inconsistently across different banks. For instance, P08~(CR,SMS,M) reported: 
\begin{quote}
\textit{``Is it a passphrase or passcode or key phrase what they need? [chuckling] I think it is slightly confusing. Although I'm experienced [\ldots], it's frustrating.''}
\end{quote}

In fact, there is an astonishing number of different terms used to denote a few different forms of credentials: membership number, customer number, username, user ID; password, passcode, passphrase; memorable word, memorable answer, memorable information etc. (cf. Table~\ref{tab:identification} and~\ref{tab:auth}). As a result, participants complained they not only had to remember the actual credentials but also what they are called. P18~(SK,CR,SMS) stated: 
\begin{quote}
\textit{``One of them was called a `passcode' and the other one was `ID', and I never knew which one is which and actually. For the cases I forgot them, it was very confusing for me because when I was to recover my credentials it was difficult to know which I forgot!''} 
\end{quote}

\descr{Providing selected characters.} Participants reported a few {\em coping} strategies when asked to enter selected characters from their credentials (e.g., the third and the sixth letter), including repeating the password (or memorable word) in their head, whispering it, counting letters on their fingers, or writing it down and counting letters on paper.

Several banks even ask for the characters to be entered out of order, which makes the process significantly harder. For P17~(CR,M,P) providing characters out of order meant more effort and time: 
\begin{quote}\textit{``If I don't even bother to remember the password, I will first have to look for it, then do the math\ldots I mean it's not that difficult but ``Oh! This one is the 5th letter'', I repeat it multiple times in my head.''}
\end{quote}
Also, the methods of entering the characters seemed to matter to our participants. P11~(CR,P) stated: 
\begin{quote}
\textit{``Actually when you are accessing it from a PC it is a box, when you access it from a mobile phone it's a drop down list. I prefer the box, it's just easier. I don't have to scroll up and down and try to find it.''} 
\end{quote}

We looked more closely at the use of drop-down lists, such as those used by Barclays (both on the desktop and mobile devices) to let users enter
selected characters of their ``memorable word.'' We initially assumed their purpose was to reduce the risks of key-loggers installed on the user's device. However, with the exception of some versions of Internet Explorer, one can still type the corresponding characters from the keyboard (i.e., upon pressing a key, the correct character from the drop-down list is selected), thus users are actually still vulnerable. 
Also note that the characters are not hidden once they are selected from the menu, which makes them more vulnerable to shoulder surfing.

\descr{Memorable answer.} Beside usernames and passwords, customers of UK banks are asked to provide a ``memorable answer'' (e.g., to a question like ``What was the name of your first pet?''). In general, participants did not report problems providing their memorable answers but three of them said they were often unsure how they had spelled them. P15~(SK,P) also complained that one of their banks was asking for their memorable answer but did not say what the question was:
\begin{quote}
\textit{``HSBC gives you prompts for the memorable information. The Lloyds one just asks you to enter the memorable information but it doesn't give you prompts about what it was.''} 
\end{quote}

Three participants were concerned with the fact that their answers were easily guessable by their relatives and friends. They questioned how memorable answers actually added security to their account. P17~(CR,M,P) reported:
\begin{quote}
\textit{``This information is everywhere [\ldots] It's just a waste of authentication, and a waste of time.''} 
\end{quote}

\descr{Assigned usernames.} The majority of UK banks assign usernames (that cannot be changed) -- which
often makes it harder for users to remember them. Participants in our studies reported two main coping strategies: eight of them said they were writing their usernames down, while seven had their browsers remember them. Five participants reported that not being able to remember their usernames prevented them from logging in from other computers than their own. P02~(SK,CR) stated: 
\begin{quote}
\textit{``For NatWest and HSBC, if I go somewhere else and I use another computer [\ldots] then I cannot remember it and I cannot even get through the first step.''}
\end{quote}

Participants also stressed they found their username to be an unrelatable string of numbers, unlike their usernames for other services which they were able to pick themselves.

\descr{Passwords.} When asked to create a password, three participants stated they reused an existing password and nine that they partially reused it. P07~(CR,P) said: 
\begin{quote}
\textit{``I use one password for everything, and I vary it a bit I add a number an exclamation mark, a hashtag, but the core password is the same for all websites.''}
\end{quote}
Seven participants said they created participants to be highly memorable (to facilitate their recall) and other seven reported writing their passwords down. P21~(SK,SMS) struggled with creating a valid password for their bank since the instructions (describing length and type of characters) did not match what the system actually accepted, driving P21 to use a trial-and-error strategy in order to generate an acceptable password.

\descr{Credential recovery.}
Five participants told us they had to reset their credentials at some point, describing the process as cumbersome. P12~(CR) reported: 
\begin{quote}
\textit{``You have a number of times you can repeat the process before they lock you out. And then you have to phone them up and go through the whole thing and get them to reset it and it takes 24 hours. So you need to be very very careful [\ldots] And I always get a bit nervous -- the last thing you want to do is be locked out from your account. It's really frustrating as well because you could mistype m instead of n.''} 
\end{quote}

Participants that went through this process reported they started to choose easier-to-remember credentials and to focus more during logins to prevent mistakes. P11~(CR,P) said they ``learned'' from their experience:
\begin{quote}
\textit{``It's quite a big deal afterwards. To reset the whole thing, you have to call them and they have to verify things\ldots I chose easier information so that I am sure I will remember it and not have to go through the process again.''} 
\end{quote}

\subsection{Mental models of authentication tokens}
During both interviews, three participants were particularly intrigued by their hardware token and they made speculations on how it was operating. 

\descr{Where do the codes come from?} Three participants were curious about how card readers generated codes that were accepted by the website and wondered if it was connected to the Internet. A participant thought that the bank and the token have a list of codes and as soon as the customer uses one of them, it gets ``crossed out'' from the list. This belief could have been influenced by the fact that banking authentication in some European countries used to rely on this system. In fact, P10~(CR,M,P) mentioned having used this system back when they lived in Germany: 
\begin{quote}
\textit{``I used to live in Germany, I really liked the way they do it there, my bank there was Deutsche Bank. It gives slips of paper where you have OTPs which are easy to replicate\ldots For me, it's convenient to have both in the office and at home. It feels safe even if someone sees it because even if they see the code, they wouldn't know which one they would be using.''}
\end{quote}

\descr{Do I need to protect the token?} Two participants told us the token was valuable and some stated they were protective of it hiding it out of plain sight. P11~(CR,P) was worried about what kind of information related to their account could be collected from the card reader if stolen: 
\begin{quote}
\textit{``With regards to this device, I mean, it's meant to be secure I would imagine, because you are supposed to be the only person that should have possession of it. But if you lose it what happens? What kind of information can be taken off it? \ldots You know relating to your account, I am not quite sure how secure that is to be honest.''}
\end{quote}

In some of the interviews, it became apparent that participants did not realise the card readers were interchangeable between banks. When asked if it would be possible to use a card reader from one bank to log in to another (which is the case for several banks), P07~(CR,P) expressed doubts:
\begin{quote}
\textit{``Well, Nationwide sent it to me so I assume it's unique for Nationwide, I don't think I can use it for other banks. [\ldots] I think the number is unique for the debit card and the PIN, so if someone stole the reader they can't access the reader, they need to slot in the credit card. It's to make sure no one else has access to my account. To get the number, it has to be my debit card and they have to know my PIN -- so it is very unlikely\ldots unless they have the credit card the PIN and the reader.''}
\end{quote}
We also encountered another misconception regarding authentication, as P12~(CR) commented on how authentication in the morning is faster than in the evening: 
\begin{quote}
\textit{``I think the process itself can be quite slow. I think this is because a the server is quite busy and because they don't have enough staff to check this kind of stuff and that can be a problem. Especially in the evenings it seems to take longer. One day I was at home, I wanted to pay council tax 10 in the morning and it went through so fast. It was incredible. Whereas in the evening there's more people on and maybe they don't have a lot of staff at the other end checking this stuff and so this takes longer and if you are in a rush this can be frustrating.''}
\end{quote}
After a couple of clarifying questions from the interviewer, it became apparent that the participant thought verification of credentials happens manually with members of staff checking the usernames, passwords, memorable answers, and OTPs customers enter online are correct.

\descr{Rituals that make users feel secure.} Throughout the interviews, participants reported on their strategies for staying secure while banking. %
P07~(CR,P) elaborated what made them secure: 
\begin{quote}
\textit{``There is always a small chance, of a virus or people can steal your bank account details. That's always a small concern and I do as much I can to prevent that. On Windows you can have a firewall, I have Avast Antivirus, they do regular virus definition checks. I have a password, it's not an easy password and it's also memorable information. I change them every 6 months and if I log in from a friend's computer or a library computer, I don't ask the computer to remember my username. I don't do that.''}
\end{quote}

\subsection{Motivation for online banking}
During the first interviews, we asked participants to discuss their preferred ways of banking. 
Out of the 21 participants, 17 reported they preferred online banking, with 11 participants appreciating its convenience,
six its ease of use, and seven the fact that online banking is less time-consuming than other forms of banking. 
Obviously, this strong preference might be correlated to the fact that we recruited participants for an online banking study.
Participants also highlighted several reasons why they prefer not to go to bank branches,
including long queues and/or distance to the nearest branch, whereas, four of them reported preferring in-person
banking due to faster/better resolution of queries and having face-to-face contact with members of staff.

\subsection{Ideal authentication}
As part of the second round of interviews, we asked participants what an ideal authentication procedure for online banking would look like --
we now analyse some of their answers.

\descr{Biometrics.} Authentication using biometrics was the most frequently proposed idea.
Some participants mentioned it already in the first interview, before we even asked this question. For example,
P06~(CR,SMS,P), after discussing their struggle with username, password, and token based authentication, told us:
\begin{quote}
\textit{``If you could be in an ideal world where you wouldn't have to use your card reader, type in your password\ldots you know I think, in a few 100 years from now you'll just put your finger on a machine and it reads your fingerprint. Today, it's slow -- you know fast is good! The faster the better.''}
\end{quote}
P17~(CR,M,P) also said they would like a camera-based biometric system, as long as the process would be fast: 
\begin{quote}
\textit{``Maybe through the camera, webcam looks at me and verifies it's me entering the five-digit code[\ldots] I would be willing to change my position and stay still until the scan finishes -- I still don't need to remember anything. [\ldots] I think 5 seconds or 10 maximum, not beyond 10 seconds -- I'd freak out -- I'd prefer to remember.''}
\end{quote}

\descr{Reducing cognitive effort.} Six participants emphasised they would like an authentication system that minimises the need to remember anything. Seven participants also emphasised they needed to concentrate to log in. They stressed that especially when tired or in a rush, they needed to stop what they were doing and focus. P14~(SMS) stressed they needed to make a conscious attention switch:
\begin{quote}
\textit{``If I am in a rush, I maybe misspell my surname or I do not enter the card number correctly. If there's any delays, it usually is one of those two. The device itself, I don't think I ever had problems with that stage, it's always been the first one. I'll have to get myself together mentally and let's say ``Focus! Whatever is in your mind, forget it.''}
\end{quote}

\descr{Reducing physical effort.} Participants expressed their desire for an authentication mechanism that would reduce the physical effort they need to make -- by avoiding the need to enter multiple pieces of information (5 mentions), or to carry additional devices with them (4).

\descr{Fast and simple.} Reducing the cognitive and physical burden would lead to faster and simpler authentication procedures. In general, participants preferred fewer steps, as some of them suggested they should be logged in based on their actions, passively, rather than actively needing to enter their credentials. For instance, P06~(CR,SMS,P), while suggesting the use of biometrics, said:
\begin{quote}
\textit{``We have our unique fingerprint no one can replicate at the moment, and you just put your finger on the screen and that's it, a one-step process.''}
\end{quote}

Also note that three participants said they preferred one bank to another because the login process was faster.
 
\descr{Portable.} Two participants also highlighted the need for ``authentication portability.'' That is, logins should be for more than just one system -- for example, when using their fingerprint to access their computer, this should already log them in to their bank and also to other accounts.

\subsection{Implicit authentication}
At the end of the second round of interviews, we also asked participants about their views on implicit authentication.
We explained, everyday English, how an implicit authentication mechanism works, that it authenticates users by constantly monitoring some aspects of their online activity and their behaviour.
We heard two main reservations about this type of authentication. First, five participants said they system would need to be highly accurate as they did not want to find themselves locked out of their machines. Second, six participants voiced privacy concerns saying they might find it intrusive and could feel uncomfortable knowing their behaviour was being tracked. They also stressed they would want to know who is storing their data and how it is being used. 
P10~(CR,M,P) stressed:
\begin{quote}
 \textit{``I could see implicit working but you'll probably run to privacy issues about that: Who's doing the software? How's the monitoring done? Who gets the information from the monitoring? blah blah blah. That would be the real issue.''}
\end{quote}

\section{Diary results}\label{sec:diary}
As stated in Section~\ref{sec:methodology}, 17 participants kept an authentication diary for an average of 11 days ($SD=4.06$), starting from the day they had their first interview. We obtained a total of 90 entries, with an average of $5.29$ entries per person (range: 1-15, $SD=3.99$). Out of 90 events, participants reported having had problems on 12 occasions ($13.3\%$). The most frequent problems were mistyped credentials with 5 occurrences and misplaced tokens with 2. The following problems were mentioned once: wrong memorable answer entered, wrong sequence of steps when using the token, a forgotten username, a frozen browser session and a slow Internet connection. 

\descr{OTP generation.} Participant could indicate their satisfaction with each login event on a five-point scale from ``very dissatisfied'' to ``very satisfied''. %
From the 90 recorded events, 11 were missing a satisfaction rating.
We compared satisfaction ratings for authentication events with and without OTP generation (30 and 60 events respectively). We found participant satisfaction was higher when they did not generate an OTP to authenticate ($U=1119$, $p<0.001$, two-tailed Mann-Whitney test).

\descr{Number of credentials required.} For each authentication event, we counted how many pieces of information participants needed to enter during the authentication process. On average, each authentication instance required $2.44$ pieces of information, ranging from 1 to 5 (e.g., 5 pieces of information were: surname, account number, last 4 digits of card, OTP, PIN for token). We found this quantity and satisfaction were negatively correlated ($\rho=-0.48, p<0.001$): the more pieces of information were needed, the lower participant satisfaction was.

\descr{Time of day.} We divided the events in four times of the day: morning (6am-12pm), afternoon (12pm-6pm), evening (6pm-12am) and night (12am-6am). We found 24 events took place in the morning, 32 in the afternoon, 30 in the evening, and 4 during the night. %
We noticed that morning events had a higher satisfaction score than events at other times of the day ($4.51$ vs. $4.07$), and found this difference statistically significant ($U=978$, $p<0.01$, two-tailed Mann-Whitney test). We also checked whether participants were more likely to encounter authentication problems in the evening (as opposed to during the rest of the day): participants experienced 8 problems out of the 30 authentication events occurring in the evening versus 4 problems out of the 60 events during the rest of the day. Note that this difference is statistically significant ($p=0.0175$, two-tailed Fisher's exact test).

\descr{Type of hardware token.} We also compared satisfaction levels for banks using two different types of hardware tokens: events with banks using a card reader had significantly higher satisfaction than events with banks using Secure Key ($U=203$, $p=0.001$, two-tailed Mann-Whitney test).
 
\section{Discussion \& Conclusion}\label{sec:conclusion}
This paper presented an in-depth user study of the usability of two-factor authentication (2FA) in the context of online banking. We focused on UK banks and their use of 2FA for website login, payment setup, and login onto smartphone apps. We conducted two series of semi-structured interviews, collected entries from an authentication diary over an average of 11 days, and analysed data both qualitatively and quantitatively.

\descr{Summary of findings.} Participants reported using 2FA without too many mistakes or lockouts, which is not very surprising as we deliberately recruited customers accustomed to 2FA and online banking.
On the other hand, we uncovered a number of issues with credentials in general and, in particular, with 
hardware tokens. The demands placed on users, such as the need to produce and remember a wide range of different credentials (often with confusing nomenclatures), or having to carry around and operate extra devices, actually shape how often, from where, and how, customers use online banking. 
Interviews, as well as the analysis of authentication diaries, showed that participants' satisfaction dropped when they had to provide multiple pieces of information and, above all, use hardware tokens.
Several users were not happy with these as they could not automate the authentication process and were disrupted from their primary task (e.g., by having to stop and think about what to do next). One participant actually switched banks to avoid having to use hardware tokens.

Several participants proposed biometric-based authentication as an ideal solution that would require less effort (nothing to carry or to remember). Implicit authentication was also a well-received idea, but participants stressed it would need to be highly accurate and respectful of their privacy. 
Participants asserted, on several occasions, they knew what kind of level of security was required for them and took measures to obtain it. If participants found that their memorable answer was easy to guess, they used a password instead. 
They also found authentication procedures to be excessive compared to offline transactions where only a four-digit PIN is needed to authorise a card payment. %

\descr{Recommendations.} We found that reducing the number of steps required for online banking authentication was crucial for our participants. For most of them, an ideal authentication process would involve fewer steps and would not require them to carry tokens. 
This highlights the need to reduce the number of steps throughout the authentication process, and to minimise the use of the hardware token. As participants have different capabilities, needs and preferences, banks should provide customers with a choice of which 2FA method they can use. For instance, they could let them optionally replace hardware tokens with codes received via SMS or generated by a smartphone app, and/or restrict their use to setting up payments above a certain threshold.

Our study also yields a few immediately actionable suggestions. In order to minimise confusion, UK banks should unify the wording for credentials (which now includes password, passcode, passphrase, memorable word, memorable information, etc.). The use of different authentication terminology by each bank was a particular burden for participants who had accounts with more than one bank (16/21 in our sample). This highlights the need for banks to not consider their authentication in isolation, but the actual tasks and contexts of use that their customers face, and how design of the mechanism impacts the lived experience of their customers.
We also argue that banks should remove features that do not add any real security but actually negatively affect the user experience by increasing disruption or time needed to authenticate. For instance, some banks, aiming to thwart key-looking attacks, force users to enter selected password characters by choosing from drop-down menus, however, this is cumbersome for many users and not really effective as one can still type the corresponding characters from the keyboard. (Also the characters are not hidden once they are selected from the menu, which makes them more vulnerable to shoulder surfing.)

\descr{Study limitations.} Our sample (21 participants) is small and consists of well-educated, relatively young individuals –- this is representative of those who currently adopt online banking, but also an `easy' user group. Given that this group of participants struggled, it is likely that older, less technology-savvy user groups are put off online banking by the lack of usability which is an accessibility issue, but also a missed commercial opportunity for banks looking to lower transaction costs. We mostly rely on self-reported data, which might be subject to exaggeration, selective memory, or social desirability. The findings from our study can be tested in a future, larger online study. Objective empirical data about success and failure, password resets, and customers leaving can only be obtained from the banks themselves, who so far have been hiding behind ``commercial sensitivity'' arguments.

\descr{Future work.} Our work paves the way to several possible follow-up studies. Besides designing a larger, confirmatory study, we also plan to conduct statistical analysis to empirically compare usability across different 2FA technologies for online banking. To this end, we are working on a privacy-preserving infrastructure geared to collect data measuring  authentication task's completion time, mistakes, lockouts, stress, frustration etc. We also hope that future research, based on our initial findings, will work on redesigning interfaces, technologies, and interactions to improve user experience and satisfaction with online banking authentication.

\descr{Acknowledgements.} Authors wish to acknowledge Granville Moore, Steven Murdoch, Iacovos Kirlappos, Ingolf Becker, Eugene Vasserman, and Greg Norcie for their helpful comments and feedback.
Kat Krol's work was supported by an EPSRC grant to the Security Science Doctoral Training Centre (grant no: EP/G037264/1). 

\bibliographystyle{IEEEtranS}
\bibliography{biblio}
\end{document}